\documentclass[conference]{IEEEtran}

\usepackage{graphicx}
\usepackage{tikz}
\usepackage{pgfplots}

\newcommand*{\affmark}[1][*]{\textsuperscript{#1}}

\ifCLASSINFOpdf
\else
\fi

\usepackage{amssymb}
\usepackage[bb=dsserif]{mathalpha}
\usepackage{amsmath}
\usepackage{cite}
\usepackage[nolist]{acronym}

\newcommand{\prob}{\mathbb{P}}
\newcommand{\generator}{\mathbf{G}}
\newcommand{\innerMarker}{\mathrm{Inn}}
\newcommand{\outerMarker}{\mathrm{Out}}
\newcommand{\generatorInner}{\mathbf{G}_\innerMarker}
\newcommand{\generatorOuter}{\mathbf{G}_\outerMarker}
\newcommand{\identity}{\mathbf{I}}
\newcommand{\parityBuild}{\mathbf{P}}

\newcommand{\parityCheck}{\mathbf{H}}
\newcommand{\codewordSize}{n}
\newcommand{\messageSize}{k}
\newcommand{\redundancySize}{r}

\newcommand{\rvmsg}{M^\messageSize}
\newcommand{\messageName}{m^\messageSize}
\newcommand{\queryIndex}{q}
\newcommand{\codeword}{c^\codewordSize}
\newcommand{\codewordGuess}{c^{\codewordSize, (\queryIndex)}}

\newcommand{\rvword}{C^\codewordSize}
\newcommand{\rvwordi}{C_i}
\newcommand{\binaryAlphabet}{\mathbb{F}_2}
\newcommand{\codebook}{\mathcal{C}}
\newcommand{\codeInner}{\mathcal{C}_\innerMarker}
\newcommand{\codeOuter}{\mathcal{C}_\outerMarker}

\newcommand{\rvchanout}{Y^\codewordSize}
\newcommand{\rvchanouti}{Y_i}
\newcommand{\rvchanoutSingleParity}{Y_{\messageSize + 1}}

\newcommand{\rveffect}{Z^\codewordSize}
\newcommand{\rveffecti}{Z_i}
\newcommand{\effect}{z^\codewordSize}
\newcommand{\effecti}{z_i}
\newcommand{\msgFlips}{z^{\messageSize}}
\newcommand{\msgFlipGuess}{z^{\messageSize, (\queryIndex)}}
\newcommand{\redFlips}{z_{\messageSize + 1}^{\codewordSize}}
\newcommand{\singleParityFlip}{z_{\messageSize + 1}}
\newcommand{\rvdemodout}{D^\codewordSize}
\newcommand{\rvdemodouti}{D_i}
\newcommand{\rvMsgDemod}{D^k}
\newcommand{\rvRedDemod}{D_{k+1}^n}

\newcommand{\chanouti}{y_i}
\newcommand{\msgPart}{Y_1^\messageSize}
\newcommand{\redPart}{Y_{k+1}^\codewordSize}
\newcommand{\decodeMsg}{\hat{M}^\messageSize}
\newcommand{\decodeEffect}{\hat{Z}^\codewordSize}
\newcommand{\listSize}{L}

\begin{acronym}
    \acro{PC}{product code}
    \acro{BCH}{Bose-Chaudhuri-Hocquenghem}
    \acro{eBCH}{extended Bose-Chaudhuri-Hocquenghem}
    \acro{GRAND}{guessing random additive noise decoding}
    
    \acro{SGRAND}{soft GRAND}
    \acro{DSGRAND}{discretized soft GRAND}
    \acro{SRGRAND}{symbol reliability GRAND}
    \acro{ORBGRAND}{ordered reliability bits GRAND}
    \acro{ORB}{ordered reliability bits}
    \acro{SOGRAND}{soft-output GRAND}
    \acro{5G}{the $5$-th generation wireless system}
    \acro{NR}{new radio}
	\acro{APP}{a-posteriori probability}
	\acro{ARQ}{automated repeat request}
	\acro{ASK}{amplitude-shift keying}
	\acro{AWGN}{additive white Gaussian noise}
	\acro{B-DMC}{binary-input discrete memoryless channel}
	\acro{BEC}{binary erasure channel}
	\acro{BER}{bit error rate}
	\acro{biAWGN}{binary-input additive white Gaussian noise}
	\acro{BLER}{block error rate}
	\acro{uBLER}{undetected block error rate}
	\acro{bpcu}{bits per channel use}
	\acro{BPSK}{binary phase-shift keying}
	\acro{BSC}{binary symmetric channel}
	\acro{BSS}{binary symmetric source}
	\acro{CDF}{cumulative distribution function}
	\acro{CRC}{cyclic redundancy check}
	\acro{DE}{density evolution}
	\acro{DMC}{discrete memoryless channel}
	\acro{DMS}{discrete memoryless source}
	\acro{BMS}{binary input memoryless symmetric}
	\acro{eMBB}{enhanced mobile broadband}
	\acro{FER}{frame error rate}
	\acro{uFER}{undetected frame error rate}
	\acro{FHT}{fast Hadamard transform}
	\acro{GF}{Galois field}
	\acro{HARQ}{hybrid automated repeat request}
	\acro{i.i.d.}{independent and identically distributed}
	\acro{LDPC}{low-density parity-check}
	\acro{GLDPC}{generalized low-density parity-check}
	\acro{LHS}{left hand side}
	\acro{LLR}{log-likelihood ratio}
	\acro{MAP}{maximum-a-posteriori}
	\acro{MC}{Monte Carlo}
	\acro{ML}{maximum-likelihood}
	\acro{PDF}{probability density function}
	\acro{PMF}{probability mass function}
	\acro{QAM}{quadrature amplitude modulation}
	\acro{QPSK}{quadrature phase-shift keying}
	\acro{RCU}{random-coding union}
	\acro{RHS}{right hand side}
	\acro{RM}{Reed-Muller}
	\acro{RV}{random variable}
	\acro{RS}{Reed–Solomon}
	\acro{SCL}{successive cancellation list}
	\acro{SE}{spectral efficiency}
	\acro{SNR}{signal-to-noise ratio}
	\acro{UB}{union bound}
	\acro{BP}{belief propagation}
	\acro{NR}{new radio}
	\acro{CA-SCL}{CRC-assisted successive cancellation list}
    \acro{SA-GCD}{single parity-check aided GCD}
	\acro{DP}{dynamic programming}
	\acro{URLLC}{ultra-reliable low-latency communication}
    \acro{GCC}{Generalized Concatenated Code}
    \acro{GCCs}{Generalized Concatenated Codes}
    \acro{MDS}{maximum distance separable}
    \acro{ORBGRAND-AI}{ORBGRAND Approximate Independence}
    \acro{BLER}{block error rate}
    \acro{crc}[CRC]{cyclic redundancy check}
    \acro{cascl}[CA-SCL]{CRC-Assisted Successive Cancellation List}
    \acro{uer}[UER]{undetected error rate}
    \acro{MDR}{misdetection rate}
    \acro{LMDR}{list misdetection rate}
    \acro{QC}{quasi-cyclic}
    \acro{SISO}{soft-input soft-output}
    \acro{VN}{variable node}
    \acro{CN}{check node}
    \acro{SI}{soft-input}
    \acro{SO}{soft-output}
    \acro{PAC}{polarization-adjusted convolutional}
    \acro{dRM}{dynamic Reed Muller}
    \acro{QPSK}{quadrature phase shift keying}
    \acro{GCD}{guessing codeword decoding}
    \acro{OSD}{ordered statistics decoding}
    \acro{SLVA}{serial list Viterbi algorithm}
    \acro{RLC}{random linear code}
    \acro{SPC}{single full parity-check}
    \acro{DARPA}{Defense Advanced Research Projects Agency}
\end{acronym}

\begin{document}

\title{Using a Single Parity-Check to Reduce the Guesswork of
Guessing Codeword Decoding}

\author{
	\IEEEauthorblockN{Joseph Griffin\affmark[1], Peihong Yuan\affmark[1], Ken R. Duffy\affmark[2], Muriel M\'edard\affmark[1]}\\
	\IEEEauthorblockA{\affmark[1]\textit{Research Laboratory of Electronics, Massachusetts Institute of Technology} \\   
	\affmark[2]\textit{College of Engineering and College of Science, Northeastern University}\\
	\texttt{\{joecg, phyuan, medard\}@mit.edu, k.duffy@northeastern.edu}}
}

\maketitle

\begin{abstract}
Guessing Codeword Decoding (GCD) is a recently proposed soft-input forward error correction decoder for arbitrary binary linear codes.  Inspired by recent proposals that leverage binary linear codebook structure to reduce the number of queries made by Guessing Random Additive Noise Decoding (GRAND), for binary linear codes that include a full-message single parity-check (SPC) bit, we show that it is possible to reduce the number of queries made by GCD by a factor of up to 2 with the greatest guesswork reduction realized at lower SNRs, without impacting decoding precision.
Codes without a full-message SPC can be modified to include one by changing a column of the generator matrix to obtain a decoding complexity advantage, and we demonstrate that this can often be done without losing decoding precision. To practically avail of the complexity advantage, a noise effect pattern generator capable of producing sequences for given Hamming weights, such as the landslide algorithm developed for ORBGRAND, is necessary.
\end{abstract}

\IEEEpeerreviewmaketitle

\section{Introduction}
\label{section:intro}
Forward error correction decoding algorithms that can function with any code construction have been a focus of research in recent years.  Such decoding strategies primarily rely on sequentially querying the impact of putative noise on hard-decision symbols. Noise effect guesswork over the entire received symbol vector, as in Guessing Random Additive Noise Decoding (GRAND) \cite{Duffy19}, enables the decoding of any code that has an efficient codebook membership test, and has been demonstrated on code structures including binary linear codes \cite{An21}, nonlinear codes \cite{cohen2023aes}, codes in higher-order fields, and codes based on length constraints \cite{ozaydin2022grand}, in both hard and soft detection settings. Soft detection informed noise effect queries on the message section of a linear code with a systematic parity check matrix forms the basis of development of Guessing Codeword Decoding (GCD) \cite{ma2024gcd,zheng2024universal}, which supports arbitrary binary linear code structures.

GRAND operates by guessing noise effects in decreasing order of likelihood based on statistical knowledge of the channel or soft information.  Each noise effect is inverted against the received signal according to the channel model, and the codebook is checked for the resulting putative transmission.  If the putative transmission is a valid codeword, because noise sequences were generated in decreasing likelihood order, then the first codeword generated in this manner is the \ac{ML} codeword. Similarly, a list of codewords generated in this way will also be ordered by decreasing likelihood \cite{abbas2021listgrand}.

GCD operates by guessing noise effects as inverted from a pre-selected information set of the codeword, e.g. the message bits of a systematic code.  The resulting putative message noise effect is re-encoded by the receiver to determine the total likelihood of the information set corresponding to the guessed noise effect.  
When the likelihood cost of a putative message noise effect outweighs the likelihood cost of the best codeword noise effect already found, guesswork can be stopped \cite{taipale1991improvement,ma2024gcd,zheng2024universal} and the most likely codeword found so far can then be output as the \ac{ML} decoding. GCD leverages the efficient encoding feature of codes with binary linear structure to reduce guesswork.

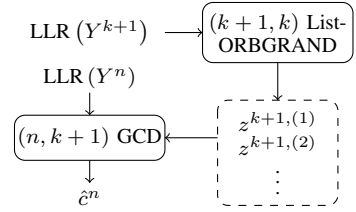
\begin{figure}[t]
	\centering
	\begin{tikzpicture}[scale=1]
    \footnotesize

\node at (-2.5,0)[align=center]{$\text{LLR}\left(Y^{k+1}\right)$}; 
\draw[->] (-1.5,0) -- (-1,0) ;

\draw[rounded corners] (-1,-0.4) rectangle (1,0.4);\node at (0,0)[align=center]{$(k+1,k)$ List-\\ORBGRAND}; 

\draw[->] (0,-0.4) -- (0,-0.9) ;

\draw[rounded corners,dashed] (-0.8,-0.9) rectangle (0.8,-2.3);\node at (0,-1.6)[align=center]{$z^{k+1,(1)}$\\$z^{k+1,(2)}$\\$\vdots$}; 

\node at (-2.5,-0.6)[align=center]{$\text{LLR}\left(Y^{n}\right)$}; 
\draw[->] (-2.5,-0.8) -- (-2.5,-1.1) ;

\draw[->] (-0.8,-1.4) -- (-1.5,-1.4) ;
\draw[rounded corners] (-3.5,-1.7) rectangle (-1.5,-1.1);\node at (-2.5,-1.4)[align=center]{$(n,k+1)$ GCD}; 

\draw[->] (-2.5,-1.7) -- (-2.5,-2.0) ;
\node at (-2.5,-2.2)[align=center]{$\hat{c}^{n}$};

\end{tikzpicture}
	\caption{Block diagram for ORBGRAND and GCD decoding of an abritrary systematic binary linear code concatenated with an SPC.}
	\label{fig:concat_diagram}
    \vspace{-18pt}
\end{figure}

This paper examines the behavior of ORBGRAND and GCD as executed on component codes of a concatenated code.  ORBGRAND is operated as a list decoder on the outer code, and the list output is used as a guesswork generator for noise sequences re-encoded to match the inner code according to the GCD scheme, see Figure \ref{fig:concat_diagram}.  The guesswork ordering from list-GRAND output causes GCD to reach its stopping condition with fewer re-encodings than guesswork produced exclusively from the information set, as in \cite{ma2024gcd}, without loss in decoding precision.  Because both of these decoders support arbitrary binary linear codes, our scheme supports any concatenated code structure.  When the reduction in GCD re-encodings exceeds the guesswork introduced by list-ORBGRAND, this joint decoder system achieves complexity gains over a comparable implementation of GCD alone.  Therefore we consider outer codes with which list-ORBGRAND introduces minimal guesswork per codeword.

Inspired by proposals that exploit the structure of binary linear codebooks to reduce GRAND guesswork \cite{rowshan2022constrained,rowshan2023lowcomplexity,rowshan2023segmented},
this paper examines the use of the joint GRAND-GCD decoder structure with binary linear codes that contain a full-message single parity-check (SPC) sub-code.  The SPC sub-code structure can be decomposed into an arbitrary linear inner code concatenated with an SPC outer code.  Exploitation of that SPC constraint, which represents an even outer code \cite{lin_error_2004,rowshan2022constrained}, requires a soft input noise effect query generator capable of solely generating patterns of given Hamming weights. Thus equipped, the list-ORBGRAND decoder can produce a new valid outer codeword with every putative noise effect guess, introducing no further guesswork beyond that of GCD alone.  The result is that low-likelihood guesswork is shifted later in the guesswork order, allowing GCD to reach its stopping condition with less total guesswork without skipping potentially likely codeword guesses.  We refer to the joint decoder system as \ac{SA-GCD}.

To identify true \ac{ML} decodings in soft detection settings, both GCD and GRAND require that putative noise effect sequences for received hard-decision bits can be generated in decreasing order of likelihood from channel \acp{LLR}, such as the one introduced in \cite{solomon20}. For practical implementation, however, approximations are necessary. In particular,  Ordered Reliability Bits GRAND (ORBGRAND) develops a guesswork generator based on a statistical approximation that reduces the ordered guesswork generation problem to an integer-partition problem.  This approximation has little impact on the \ac{BLER} \cite{duffy22ORBGRAND} and results in a decoder that is almost capacity achieving \cite{Liuetal23}, but which is readily implementable in circuits, e.g. \cite{condo2021_highperformance,abbas2021orbgrand,condo2021fixed,ORBGRAND_low_power}. Moreover, the error pattern generator introduced in \cite{duffy22ORBGRAND} and implemented in \cite{ORBGRAND_low_power, Riaz24} has the desired property of being able to create sequences of given Hamming weight. We refer to this generator as the landslide generator. Previous work on GCD have directly employed ORBGRAND's query order without availing of this configurable Hamming weight property, e.g. \cite{ma2024gcd, duffy2024soft}, and we refer to a direct GCD implementation based on the landslide generator as ORB-GCD.

Amongst other developments, the even code proposal explored in \cite{rowshan2022constrained, rowshan2023segmented, rowshan2023lowcomplexity} describes a strategy for accelerating ORBGRAND with an even code: for any code in which all valid codewords have even parity, noise sequences with parity that do not match that of the hard-decision received bits can be skipped. The landslide generator can be configured to skip generating putative noise effects with Hamming weights that fail this criterion. An SPC code always has this property, as the sole parity constraint produces even parity across the codeword.

Avoiding unnecessary guesswork would be desirable for complexity reduction of GCD.  The \ac{SLVA} has been explored for GRAND \cite{PC_GRAND} to accelerate decoding by skipping guesswork.  One disadvantage of the \ac{SLVA} is that it requires extra computation to determine what guesswork to skip.  By comparison, the guesswork skipping proposed in \cite{rowshan2022constrained} requires a single parity computation on the received hard-decision bits, with no other online computation. As a result, in a circuits implementation, the energy and latency for ORBGRAND with even codes (and therefore \ac{SA-GCD}) would likely be lower than for an \ac{SLVA}-driven technique.

\Ac{SA-GCD} reduces GCD guesswork by up to a factor of 2 at lower \acp{SNR}.  This development can be understood as an exploitation of properties of the landslide generator \cite{duffy2022_ordered,ORBGRAND_low_power} and the even code observation in \cite{rowshan2022constrained} as applied to GCD \cite{ma2024gcd}, but with a twist where further understanding comes from a perspective in which the code is jointly decoded by ORBGRAND and GCD on components of a concatenated code.

This paper is organized as follows: Section \ref{section:problem_setting} describes the setting and code structure in which \ac{SA-GCD} offers gains over ORB-GCD.  Section \ref{section:proposal_description} describes \ac{SA-GCD} and explains why the complexity gains arise.  Section \ref{section:results} provides numerical results for experiments in which \ac{SA-GCD} is directly compared to the ORB-GCD.  Section \ref{section:conclusion} summarizes this work.

\section{Problem Setting}
\label{section:problem_setting}
We consider a sender and receiver connected by a memoryless channel with binary input.  The sender has a random  $\messageSize$-bit binary message vector $\rvmsg$ distributed uniformly over $\binaryAlphabet^\messageSize$ with realized value $\messageName$.  Lowercase letters are used to denote realizations of random variables.  Let $\mathbf{T}$ denote an arbitrary full-rank element of $\mathbb{F}_2^{k\times k}$.  The sender and receiver first agree on a binary linear code represented by generator matrix $\generator = \mathbf{T}[\identity~\parityBuild] \in \binaryAlphabet^{\messageSize \times \codewordSize}$ with $\redundancySize = \codewordSize - \messageSize$ and its systematic-form parity-check matrix $\parityCheck = [-\parityBuild^\mathrm{T}~\identity] \in \binaryAlphabet^{\redundancySize \times \codewordSize}$, where $\identity$ is the identity matrix and $\parityBuild \in \binaryAlphabet^{\messageSize \times \redundancySize}$.  The message $\messageName$ is encoded using $\generator$ to form the codeword $\codeword = \messageName \generator$, an element of the codebook $\codebook$, the row space of $\generator$.

The codeword is transmitted over the channel, and the receiver records the random, continuous-valued received signal $\rvchanout$.  Let $f_{Y|C}(\cdot | \cdot)$ denote the probability density function of $Y$ given $C$.  Each symbol $\rvchanouti$ then has an associated \ac{LLR} defined as
\begin{align*}
    \mathrm{LLR}(\rvchanouti) = \log{\frac{f_{Y|C}(\rvchanouti|0)}{f_{Y|C}(\rvchanouti|1)}}
    .
\end{align*}
The demodulation of $\rvchanouti$ is then $\rvdemodouti = (\mathrm{sign}(\mathrm{LLR}(\rvchanouti)) + 1) / 2 \in \binaryAlphabet^\codewordSize$, an estimate of $\rvwordi$.  The noise effect $\rveffect$ is defined to be $\rveffect = \rvdemodout \ominus \rvword$.

As $\rveffecti = 1$ if and only if $\rvdemodouti \neq \rvwordi$, we can use the \ac{LLR} definition to evaluate the probability that each demodulated symbol is incorrect based on the channel output:
\begin{align*}
    \prob(\rvdemodouti \neq \rvwordi) = \prob(\rveffecti = 1) = \frac{e^{-|\mathrm{LLR}(\rvchanouti)|}}{1 + e^{-|\mathrm{LLR}(\rvchanouti)|}} \in \left.\left(0, \frac{1}{2}\right]\right.
    .
\end{align*}
The \textit{a priori} \ac{BER} $\prob(\rveffecti = 1)$ decreases monotonically with $|\mathrm{LLR}(\rvchanouti)|$, so $|\mathrm{LLR}(\rvchanouti)|$ is called the reliability of $\rvdemodouti$.  The decoder assumes the elements of $\rvchanout$ are distributed independently, so $|\mathrm{LLR}(\chanouti)|$ can be used to produce noise sequences $\effect$ in decreasing order of $\prob(\rveffect = \effect)$ \cite{duffy22ORBGRAND}.  The independence assumption allows each sequence probability to be generated from the symbol probabilities:
\begin{align*}
    \prob(\rveffect = \effect) = \prod_{i = 1}^{n}\prob(\rveffecti = \effecti)
    .
\end{align*}
For a binary linear code, the \ac{ML} decoding of $\rvchanout$ is found by determining the \ac{ML} $\effect$ for which $(\rvdemodout \ominus \effect)\parityCheck^T = 0$ \cite{Duffy19}.

We call $\msgPart$ and $\redPart$ the message portion and redundant portion of the received signal, respectively, with hard-decision vectors $\rvMsgDemod$ and $\rvRedDemod$.  Let $\decodeMsg$ denote the message portion of the decoder output, and let $\decodeEffect = \rvdemodout \ominus \decodeMsg \generator$ denote the $\effect$ in the final query of a GRAND decoder.  When discussing GCD, it is also useful to discuss message part $\msgFlips$ and redundancy part $\redFlips$ separately, with $\msgFlipGuess$ being the $\msgFlips$ value guessed on the $\queryIndex^\mathrm{th}$ query.  Note that a given message noise effect $\msgFlips$ has a reliability cost given by
\begin{align*}
    \mathrm{cost}(\msgFlips) = \sum_{i = 1}^{k}\effecti |\mathrm{LLR}(\rvchanouti)|.
\end{align*}
The total reliability cost of a $\effect$ guess is the sum of $\mathrm{cost}(\msgFlips)$ and $\mathrm{cost}(\redFlips)$, and the \ac{ML} $\effect$ has minimal reliability cost.  Finally, we denote the codeword guessed by GCD on the $\queryIndex^\mathrm{th}$ guess as $\codewordGuess = \rvdemodout \ominus [\msgFlipGuess\quad\rvdemodout\parityCheck^\mathrm{T}\oplus\msgFlipGuess\parityBuild]$.

\begin{figure*}[t]
	\centering
	\begin{tikzpicture}[scale=1]
    \footnotesize

\node at (0,3)[align=center] {
$\mathbf{G}=\begin{bmatrix}
1  &   0  &   1  &   0  &   1  &   1  &   0\\
1  &   1  &   1  &   1  &   0  &   0  &   0\\
0  &   0  &   1  &   1  &   0  &   1  &   1
\end{bmatrix}$ \\
$\qquad=\begin{bmatrix}
1  &   0  &   1 \\
1  &   1  &   1 \\
0  &   0  &   1 
\end{bmatrix}\begin{bmatrix}
\mathbf{I}_3 & \begin{matrix}
 1 \\
 1 \\
 1 
\end{matrix}\end{bmatrix}\begin{bmatrix}
\mathbf{I}_4 & \begin{matrix}
 0 & 0 & 1\\
 0 & 1 & 0\\
 1 & 1 & 1\\
 1 & 0 & 0
\end{matrix}\end{bmatrix}$
};

\node at (9,3)[align=center] {\begin{tabular}{ c|ccccccc } 
\hline
$C^n$ & $0$ & $1$ & $0$ & $1$ & $1$ & $1$ & $0$\\ 
  \hline
$\text{LLR}\left(Y^n\right)$ & $-0.1$ & $+0.4$ & $+3.0$ & $-5.0$ & $-1.0$ & $-2.0$ & $+2.0$\\ 
 \hline
 \hline
 $D^n$ & $1$ & $0$ & $0$ & $1$ & $1$ & $1$ & $0$\\ 
  \hline
$\left|\text{LLR}\left(Y^n\right)\right|$ & $0.1$ & $0.4$ & $3.0$ & $5.0$ & $1.0$ & $2.0$ & $2.0$\\ 
\hline

\end{tabular}
};

\node at (9.5,1) {SA-GCD};
\node at (7.2,0.5) {$q$};
\node at (8,0.5) {$z^{k+1}$};
\node at (9.8,0.5) {$\text{cost}\left(z^{k+1}\right)$};
\node at (11.4,0.5) {$\text{cost}\left(z^n\right)$};

\node at (7.2,0) {$1$};
\draw[draw=none,fill=red!30] (7.6,-0.2) rectangle (8.4,0.2);
\node at (8,0) {$0000$};
\draw[draw=none,fill=gray!30] (8.4,-0.2) rectangle (9,0.2);
\node at (8.7,0) {$011$};
\node at (9.8,0) {$0.0$};
\node at (11.4,0) {$4.0$};

\node at (7.2,-0.5) {$2$};
\draw[draw=none,fill=red!30] (7.6,-0.7) rectangle (8.4,-0.3);
\node at (8,-0.5) {$1100$};
\draw[draw=none,fill=gray!30] (8.4,-0.7) rectangle (9,-0.3);
\node at (8.7,-0.5) {$000$};
\node at (9.8,-0.5) {$0.5$};
\node at (11.4,-0.5) {$0.5$};

\node at (7.2,-1) {$3$};
\draw[draw=none,fill=red!30] (7.6,-1.2) rectangle (8.4,-0.8);
\node at (8,-1) {$1010$};
\draw[draw=none,fill=gray!30] (8.4,-1.2) rectangle (9,-0.8);
\node at (8.7,-1) {$101$};
\node at (9.8,-1) {$3.1$};
\node at (11.4,-1) {$ $};

\node at (7.2,-1.5) {$4$};
\draw[draw=none,fill=red!30] (7.6,-1.7) rectangle (8.4,-1.3);
\node at (8,-1.5) {$0110$};
\draw[draw=none,fill=gray!30] (8.4,-1.7) rectangle (9,-1.3);
\node at (8.7,-1.5) {$110$};
\node at (9.8,-1.5) {$3.4$};
\node at (11.4,-1.5) {$ $};

\node at (7.2,-2) {$5$};
\draw[draw=none,fill=red!30] (7.6,-2.2) rectangle (8.4,-1.8);
\node at (8,-2) {$1001$};
\draw[draw=none,fill=gray!30] (8.4,-2.2) rectangle (9,-1.8);
\node at (8.7,-2) {$110$};
\node at (9.8,-2) {$5.1$};
\node at (11.4,-2) {$ $};

\draw[red,dashed,thick] (7,-0.75) rectangle (11.5,-0.75);

\node at (7.2,-2.5) {$6$};
\draw[draw=none,fill=red!30] (7.6,-2.7) rectangle (8.4,-2.3);
\node at (8,-2.5) {$0101$};
\draw[draw=none,fill=gray!30] (8.4,-2.7) rectangle (9,-2.3);
\node at (8.7,-2.5) {$101$};
\node at (9.8,-2.5) {$5.4$};
\node at (11.4,-2.5) {$ $};

\node at (7.2,-3) {$7$};
\draw[draw=none,fill=red!30] (7.6,-3.2) rectangle (8.4,-2.8);
\node at (8,-3) {$0011$};
\draw[draw=none,fill=gray!30] (8.4,-3.2) rectangle (9,-2.8);
\node at (8.7,-3) {$000$};
\node at (9.8,-3) {$8.0$};
\node at (11.4,-3) {$ $};

\node at (7.2,-3.5) {$8$};
\draw[draw=none,fill=red!30] (7.6,-3.7) rectangle (8.4,-3.3);
\node at (8,-3.5) {$1111$};
\draw[draw=none,fill=gray!30] (8.4,-3.7) rectangle (9,-3.3);
\node at (8.7,-3.5) {$011$};
\node at (9.8,-3.5) {$8.5$};
\node at (11.4,-3.5) {$ $};

\node at (1.5,1) {ORB-GCD};
\node at (-0.8,0.5) {$q$};
\node at (0,0.5) {$z^{k}$};
\node at (1.8,0.5) {$\text{cost}\left(z^k\right)$};
\node at (3.1,0.5) {$\text{cost}\left(z^n\right)$};

\node at (-0.8,0) {$1$};
\draw[draw=none,fill=red!30] (-0.4,-0.2) rectangle (0.2,0.2);
\node at (-0.1,0) {$000$};
\draw[draw=none,fill=gray!30] (0.2,-0.2) rectangle (1,0.2);
\node at (0.6,0) {$0011$};
\node at (1.8,0) {$0.0$};
\node at (3.1,0) {$4.0$};

\node at (-0.8,-0.5) {$2$};
\draw[draw=none,fill=red!30] (-0.4,-0.7) rectangle (0.2,-0.3);
\node at (-0.1,-0.5) {$100$};
\draw[draw=none,fill=gray!30] (0.2,-0.7) rectangle (1,-0.3);
\node at (0.6,-0.5) {$1110$};
\node at (1.8,-0.5) {$0.1$};
\node at (3.1,-0.5) {$8.1$};

\node at (-0.8,-1) {$3$};
\draw[draw=none,fill=red!30] (-0.4,-1.2) rectangle (0.2,-0.8);
\node at (-0.1,-1) {$010$};
\draw[draw=none,fill=gray!30] (0.2,-1.2) rectangle (1,-0.8);
\node at (0.6,-1) {$1101$};
\node at (1.8,-1) {$0.4$};
\node at (3.1,-1) {$8.4$};

\node at (-0.8,-1.5) {$4$};
\draw[draw=none,fill=red!30] (-0.4,-1.7) rectangle (0.2,-1.3);
\node at (-0.1,-1.5) {$110$};
\draw[draw=none,fill=gray!30] (0.2,-1.7) rectangle (1,-1.3);
\node at (0.6,-1.5) {$0000$};
\node at (1.8,-1.5) {$0.5$};
\node at (3.1,-1.5) {$0.5$};

\node at (-0.8,-2) {$5$};
\draw[draw=none,fill=red!30] (-0.4,-2.2) rectangle (0.2,-1.8);
\node at (-0.1,-2) {$001$};
\draw[draw=none,fill=gray!30] (0.2,-2.2) rectangle (1,-1.8);
\node at (0.6,-2) {$1000$};
\node at (1.8,-2) {$3.0$};
\node at (3.1,-2) {$ $};

\draw[red,dashed,thick] (-1,-1.75) rectangle (3.5,-1.75);

\node at (-0.8,-2.5) {$6$};
\draw[draw=none,fill=red!30] (-0.4,-2.7) rectangle (0.2,-2.3);
\node at (-0.1,-2.5) {$101$};
\draw[draw=none,fill=gray!30] (0.2,-2.7) rectangle (1,-2.3);
\node at (0.6,-2.5) {$0101$};
\node at (1.8,-2.5) {$3.1$};
\node at (3.1,-2.5) {$ $};

\node at (-0.8,-3) {$7$};
\draw[draw=none,fill=red!30] (-0.4,-3.2) rectangle (0.2,-2.8);
\node at (-0.1,-3) {$011$};
\draw[draw=none,fill=gray!30] (0.2,-3.2) rectangle (1,-2.8);
\node at (0.6,-3) {$0110$};
\node at (1.8,-3) {$3.4$};
\node at (3.1,-3) {$ $};

\node at (-0.8,-3.5) {$8$};
\draw[draw=none,fill=red!30] (-0.4,-3.7) rectangle (0.2,-3.3);
\node at (-0.1,-3.5) {$111$};
\draw[draw=none,fill=gray!30] (0.2,-3.7) rectangle (1,-3.3);
\node at (0.6,-3.5) {$1011$};
\node at (1.8,-3.5) {$3.5$};
\node at (3.1,-3.5) {$ $};

\draw[thick,<->,blue] (3.7,0) -- (6.7,0);
\draw[thick,<->,red] (3.7,-0.5) -- (6.7,-2);
\draw[thick,<->,red] (3.7,-1) -- (6.7,-2.5);
\draw[thick,<->,blue] (3.7,-1.5) -- (6.7,-0.5);
\draw[thick,<->,red] (3.7,-2) -- (6.7,-3);
\draw[thick,<->,blue] (3.7,-2.5) -- (6.7,-1.0);
\draw[thick,<->,blue] (3.7,-3) -- (6.7,-1.5);
\draw[thick,<->,red] (3.7,-3.5) -- (6.7,-3.5);

\end{tikzpicture}
	\caption{Example ORB-GCD behavior vs. SA-GCD behavior for a noise realization on the $(n, k) = (7, 3)$ simplex code, with the GCD stopping condition indicated by the red line.  The reliability of $Y_4$ is high, so SA-GCD pushes queries $\msgFlips$ yielding $z_{k+1} = 1$ much later in the guessing order.  In this example, queries with $z_{k+1} = 1$ are pushed later than the stopping condition, reducing total guesswork from 4 re-encoded sequences to 2.  Note that if $|\mathrm{LLR}(Y_4)| = 0$, the guesswork of SA-GCD is identical to the guesswork of ORB-GCD because SA-GCD has no more information for guesswork ordering than ORB-GCD.}
	\label{fig:miniexample}
    \vspace{-18pt}
\end{figure*}
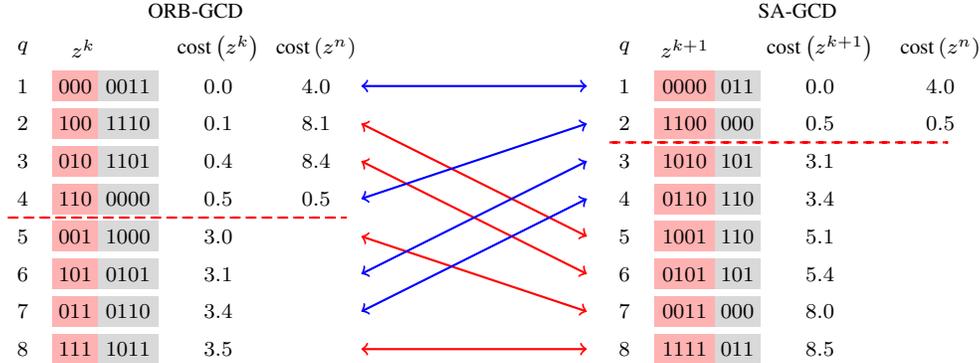

\subsection{Codebook Structure for Fast Guesswork}
\label{fast_guesswork_section}
To enable our accelerated guessing strategy, assume that the $k\times n$ generator matrix $\generator$ is constructed as a concatenation of inner and outer code generator matrices, with the outer code being an SPC code, or equivalently a $(\messageSize+1, \messageSize)$ even code.  Then $\generator$ will take the following form:
\begin{align*}
\generator = \mathbf{T}[\identity_\messageSize ~ \mathbb{1}][\identity_{\messageSize+1} ~ \parityBuild_\mathrm{Inn}],
\end{align*}
where  $\identity_\messageSize$ denotes the $\messageSize \times \messageSize$ identity matrix, $\mathbb{1}$ denotes a column vector of ones of length $\messageSize$, and $\parityBuild_\mathrm{Inn}$ denotes an arbitrary element of $\binaryAlphabet^{(\messageSize+1) \times (n-\messageSize - 1)}$. It is useful to interpret this generator as the concatenation of a potentially nonsystematic SPC outer code, $\generatorOuter=\mathbf{T}[\identity_\messageSize ~ \mathbb{1}]$,
with an inner code $\generatorInner=[\identity_{\messageSize + 1} ~ \parityBuild_\mathrm{Inn}]$. A well-studied class of codes decomposable in this way is the set of simplex codes, each of which is the dual of a Hamming code, see Figure \ref{fig:miniexample} for an example. We refer to the bit at index $\messageSize + 1$ as the SPC bit. The SPC bit represents the parity constraint that ORB-GCD ignores during guesswork generation, but that \ac{SA-GCD} exploits during guesswork generation.

A given code can be decoded faster with \ac{SA-GCD} than ORB-GCD if the generator $\generator$ can be decomposed with an appropriate outer code.  Let $\generator_i$ denote the $i^\mathrm{th}$ column of $\generator$.  \Ac{SA-GCD} can be used if there exists a set of columns $\left\{\generator_{i_j}\right\}_{j=1}^{k+1}$ such that $\dim\left(\left\{\generator_{i_j}\right\}\right) = k$ and 
\begin{equation}
\sum_{j=1}^{k}\generator_{i_j} \equiv \generator_{i_{k+1}},~\mathrm{mod}~2
\label{eq:SA_condn}.
\end{equation}
Certain forms of $\generator$ and $\parityCheck$ can be easily tested for this property.  If a systematic form of $\generator$ has $\mathbb{1}$ as a column, \ac{SA-GCD} can be used.  Similarly, \ac{SA-GCD} is applicable if $\parityCheck$ (systematic or otherwise) has any row with exactly $\messageSize+1$ nonzero elements.  In general, however, testing a code for viability of \ac{SA-GCD} requires an exhaustive search over columns of $\generator$ for an appropriate set $\left\{\generator_{i_j}\right\}$.

\section{Decoding}
\label{section:proposal_description}
With $\generator = \generatorOuter \generatorInner$, $\codeOuter$ is an SPC code that possesses the even code property.  Therefore, producing a list of given size $\listSize$ with list-ORBGRAND for $\codeOuter$ as proposed in \cite{rowshan2022constrained} requires as little as half the guesswork of ORBGRAND for an arbitrary code.  Because ORBGRAND performs guesswork in approximately \ac{ML} order, the outer code list is also approximately ordered by likelihood.   
The landslide generator can also be configured for even codes, so list-ORBGRAND on an SPC outer code is a suitable noise effect generator for GCD.

Running GCD on the inner codeword would require a list of candidate message noise effects $z^{\messageSize + 1}$ for $\codeInner$ in order of increasing cost.  Clearly list-ORBGRAND executed on $\codeOuter$ produces an appropriate list with approximately the correct order, and the only elements missing from the list will be the elements failing the outer code parity-check, which ORB-GCD executed on $\codebook$ would skip during re-encoding.  ORB-GCD executed on the full code $\codebook$ would produce $\msgFlips$ in approximately increasing cost order, but the corresponding sequence $\mathrm{cost}(z^n)$ would not be monotonic.  List-ORBGRAND produces a sequence of valid $z^{\messageSize + 1}$ corresponding to the same set of $\msgFlips$ that ORB-GCD would produce but, critically, in a different order. Namely, any $\msgFlips$ with parity mismatching $\rvMsgDemod$ would have a higher cost due to $\singleParityFlip = 1$ and would appear later in the list-ORBGRAND output than in the ORB-GCD generator output.  For example, Figure \ref{fig:miniexample} shows how SA-GCD reorders guesswork for the LLRs of a codeword from the $(n, k) = (7, 3)$ simplex code, resulting in a net reduction of guesswork by a factor of 2.

GCD reaches its stopping condition when $\mathrm{cost}(\msgFlips)>\min_q \mathrm{cost}(z^{n, (q)})$.  Therefore, reaching the stopping condition quickly requires the guesswork generator to guess $\msgFlips$ that achieves low $\mathrm{cost}(z^n)$, then jump to guessing high-cost $\msgFlips$ without missing any $\msgFlips$ for even lower $\mathrm{cost}(z^n)$.  The key to \ac{SA-GCD}'s complexity reduction without \ac{BLER} degradation is the following: List-ORBGRAND takes $\singleParityFlip$ into account when producing $\msgFlips$, so the only $\msgFlips$ that are pushed later in the guessing order are the sequences with low $\mathrm{cost}(\msgFlips)$ and high $\mathrm{cost}(z_{k+1})$.  Such guesswork naturally has high $\mathrm{cost}(z^n)$ and does not help GCD find the \ac{ML} codeword, so it does not affect the \ac{BLER}.  It also does not help GCD reach the stopping condition early, so shifting such $\msgFlips$ later in the guessing order does cut down on total guesswork.

In general, GCD guesswork spans a sequence of codewords that are not ordered according to likelihood.  Many $\msgFlips$ are encoded to no effect on the GCD state, which records the most likely codeword already guessed.  The list-ORBGRAND generator of \ac{SA-GCD} partially orders that guesswork, reducing the complexity.  The most general form of this technique would use list-ORBGRAND for an arbitrary outer code, but using an SPC outer code keeps list-ORBGRAND from guessing invalid outer codewords.

The \ac{SNR} range at which \ac{SA-GCD} pushes the largest proportion of guesswork after the stopping condition is where the stopping condition occurs after a significant number of guesses, but early enough that the extra information from the SPC bit can push guesses later in the order than the stopping condition.  That is, where the \ac{SNR} is high and the zero noise guess is usually correct, \ac{SA-GCD} saves only a small percentage of guesswork.  The stopping condition occurs early, so only a few guesses are saved by \ac{SA-GCD} over ORB-GCD.  The guesswork savings proportion grows quickly as \ac{SNR} decreases because the percentage of guesses for the inner code with both odd and even parity expands very rapidly.

In practice, \ac{SA-GCD} realizes guesswork reduction from ORB-GCD that is slightly less than a factor of 2.  The SPC outer code provides information to list-ORBGRAND that allows certain codewords to be pushed later in the guesswork list, but the information provided by the SPC bit is not perfect.  Figure \ref{fig:miniexample} shows a large guesswork reduction because $|\mathrm{LLR}(Y_{k+1})| > |\mathrm{LLR}(Y_{i})|$ for all $i\in\{1,\ldots,k\}$ in that example.  If $|\mathrm{LLR}(\rvchanoutSingleParity)|$ is close to zero, list-ORBGRAND will begin guesswork including $\singleParityFlip = 1$ quite early and will produce $\msgFlips$ with either parity.  In this case, fewer guesses are pushed later than the GCD stopping condition, and less work is saved.  With a larger message length $\messageSize$, it is less probable that $\min_{i\in\{1,\ldots,k+1\}}|\mathrm{LLR}(Y_{i})|=|\mathrm{LLR}(Y_{k+1})|$, so it is more probable that some guesswork is saved.  Therefore, larger $\messageSize$ sees a greater proportion of guesswork reduction, as can be observed by comparing Figures \ref{fig:overwork_RLC_performance} and \ref{fig:overwork_eBCH_performance}.

\section{Empirical Results}
\label{section:results}
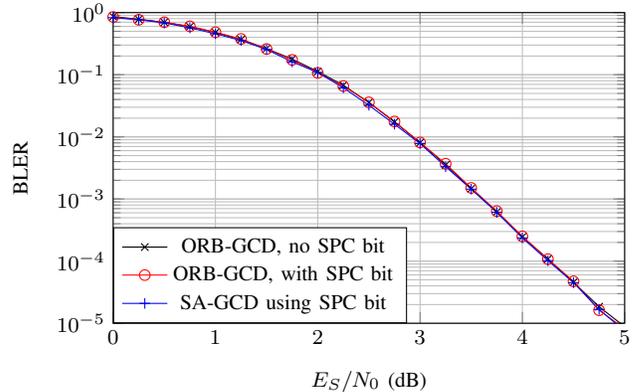
\begin{figure}[t]
\centering
\footnotesize
\begin{tikzpicture}[scale=1]
\begin{semilogyaxis}[
legend style={at={(0, 0)},anchor= south west},
ymin=1e-5,
ymax=1,
width=3.3in,
height=2.25in,
grid=both,
xmin = 0,
xmax = 5,
xlabel = {$E_S / N_0$ (dB)},
ylabel = {BLER},
]

\addplot[black, mark = x]
table{x y
0 0.86347
0.25 0.78199
0.5 0.69796
0.75 0.5954
1 0.47925
1.25 0.3683
1.5 0.26325
1.75 0.17853
2 0.11281
2.25 0.067701
2.5 0.035736
2.75 0.017345
3 0.0080758
3.25 0.0035328
3.5 0.0014648
3.75 0.0006157
4 0.00024498
4.25 0.0001084
4.5 4.57e-05
4.75 1.85e-05
5 8.8e-06
5.25 1.9e-06
5.5 1.4e-06
5.75 2e-07
6 1e-07
6.25 1e-07
6.5 0
6.75 0
7 0
7.25 0
7.5 0
7.75 0
8 0
};\addlegendentry{ORB-GCD, no SPC bit}

\addplot[red, mark = o]
table{x y
0 0.85092
0.25 0.766
0.5 0.69963
0.75 0.6022
1 0.48275
1.25 0.37681
1.5 0.25938
1.75 0.17327
2 0.10717
2.25 0.065678
2.5 0.035966
2.75 0.01768
3 0.0081273
3.25 0.0036968
3.5 0.0015073
3.75 0.00063868
4 0.00025052
4.25 0.0001076
4.5 4.73e-05
4.75 1.62e-05
5 8.2e-06
5.25 2.2e-06
5.5 1.5e-06
5.75 3e-07
6 4e-07
6.25 0
6.5 0
6.75 0
7 0
7.25 0
7.5 0
7.75 0
8 0
};\addlegendentry{ORB-GCD, with SPC bit}

\addplot[blue, mark = +]
table{x y
0 0.8284
0.25 0.77002
0.5 0.68279
0.75 0.56845
1 0.45651
1.25 0.35549
1.5 0.25511
1.75 0.16096
2 0.10907
2.25 0.061657
2.5 0.03222
2.75 0.015969
3 0.0078046
3.25 0.0033093
3.5 0.0014394
3.75 0.0006103
4 0.00023655
4.25 0.0001019
4.5 4.5e-05
4.75 1.67e-05
5 7.5e-06
5.25 2.2e-06
5.5 1.2e-06
5.75 4e-07
6 2e-07
6.25 0
6.5 0
6.75 0
7 0
7.25 0
7.5 0
7.75 0
8 0
};\addlegendentry{\Ac{SA-GCD} using SPC bit}

\end{semilogyaxis}
\end{tikzpicture}
\vspace{-24pt}
\caption{For $(128, 104)$ SPC-aided \acp{RLC}, \ac{SA-GCD} and ORB-GCD yield identical \acp{BLER}.  \Acp{RLC} without an SPC bit performed slightly worse than SPC-aided \acp{RLC} because unconstrained \acp{RLC} do not guarantee that every message bit will be protected by a parity bit.  The SPC bit in the modified \acp{RLC} always protects every bit of the message.}
\label{fig:err_rate_RLC_performance}
\vspace{-18pt}
\end{figure}

\begin{figure}[t]
\centering
\footnotesize
\begin{tikzpicture}[scale=1]
\begin{axis}[
legend style = {at = {(0, 0)}, anchor = south west},
ymin = 0,
ymax = 0.5,
width = 3.3in,
height = 1.75in,
ytick={0, 0.1, 0.2, 0.3, 0.4, 0.5},
grid = both,
xmin = 0,
xmax = 8,
xlabel = {$E_S / N_0$ (dB)},
ylabel = {Guess Count Reduction Proportion},
xtick={0, 2, ..., 8}
]

\addplot[blue, mark = +]
table{x y
0 0.26394
0.25 0.26628
0.5 0.27362
0.75 0.2822
1 0.2814
1.25 0.2973
1.5 0.29155
1.75 0.315
2 0.33684
2.25 0.35352
2.5 0.36339
2.75 0.38926
3 0.39945
3.25 0.41799
3.5 0.42733
3.75 0.43605
4 0.4519
4.25 0.45582
4.5 0.46054
4.75 0.46038
5 0.44974
5.25 0.42801
5.5 0.38768
5.75 0.33044
6 0.26749
6.25 0.20799
6.5 0.15688
6.75 0.11583
7 0.08434
7.25 0.060426
7.5 0.0425
7.75 0.029583
8 0.020126
};\addlegendentry{Relative \ac{SA-GCD} Savings}

\end{axis}
\end{tikzpicture}
\vspace{-24pt}
\caption{Guesswork savings of \ac{SA-GCD} is shown for $(128, 104)$ \acp{RLC} as a proportion of total guesswork for ORB-GCD.  The highest guesswork proportion is saved at symbol \acp{SNR} between 3 and 5 dB, where decoding involves significant guesswork but the decoder's error rate is still below $10^{-2}$.}
\label{fig:overwork_RLC_performance}
\vspace{-18pt}
\end{figure}
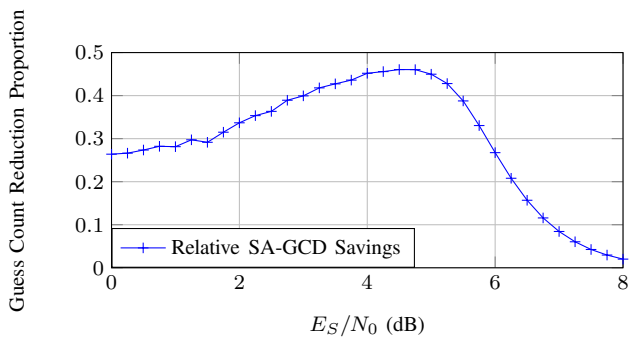

A complex-valued \ac{AWGN} channel with \ac{BPSK} input was used to demonstrate the performance of \ac{SA-GCD}.  Comparison simulations were done with \acp{RLC}, modified extended \ac{BCH} codes, and modified CRC codes.  The \ac{RLC} simulations were completed with code dimensions $(\codewordSize, \messageSize) = (128, 104)$.  Figures \ref{fig:err_rate_RLC_performance}, \ref{fig:guesswork_RLC_performance}, and \ref{fig:overwork_RLC_performance} compare \ac{SA-GCD} and ORB-GCD as executed on $\codebook$ with $\codeInner$ constructed as an \ac{RLC}.  Mimicking Shannon's original proof \cite{shannon1948}, a new code $\codebook$ was generated for each channel use to yield the provided results, and the hardware described in \cite{reuther2018interactive} carried out the simulations.

\begin{figure}[t]
\centering
\footnotesize
\begin{tikzpicture}[scale=1]
\begin{semilogyaxis}[
legend style={at={(0, 0)}, anchor= south west},
ymin=1e-4,
ymax=1e4,
width=3.3in,
height=2.0in,
ytick={1e-4, 1e-2, 1e0, 1e2, 1e4},
grid=both,
xmin = 0,
xmax = 8,
xlabel = {$E_S / N_0$ (dB)},
ylabel = {Average Guesswork Saved per Message Bit},
xtick={0, 2, ..., 8}
]

\addplot[blue, mark = +]
table{x y
0 1471.7
0.25 1403.2
0.5 1294.7
0.75 1178.9
1 993.23
1.25 819.51
1.5 632.8
1.75 448.29
2 322.37
2.25 204.04
2.5 119.94
2.75 65.681
3 34.028
3.25 16.852
3.5 7.4945
3.75 3.2935
4 1.3973
4.25 0.60552
4.5 0.26451
4.75 0.11951
5 0.05473
5.25 0.027576
5.5 0.013728
5.75 0.0076119
6 0.0045885
6.25 0.0029226
6.5 0.0019397
6.75 0.0013149
7 0.00090608
7.25 0.00062581
7.5 0.00042937
7.75 0.000294
8 0.00019777
};\addlegendentry{\Ac{SA-GCD} savings over basic ORB-GCD}

\end{semilogyaxis}

\end{tikzpicture}
\vspace{-24pt}
\caption{Average saved guesswork per bit for a $(128, 104)$ \ac{RLC} is shown for \ac{SA-GCD}.  Guesswork savings is always positive, as \ac{SA-GCD} consistently performs less guesswork on average than ORB-GCD.  At symbol \acp{SNR} between 2 and 3 dB, where decoder error rate is below $10^{-1}$, \ac{SA-GCD} typically saves around 100 guesses per message bit.}
\label{fig:guesswork_RLC_performance}

\end{figure}
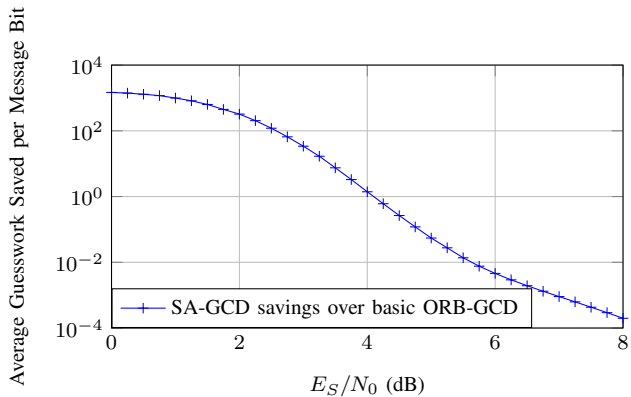

\begin{figure}[t]
\centering
\footnotesize
\begin{tikzpicture}[scale=1]
\begin{axis}[
legend style = {at = {(0, 0)}, anchor = south west},
ymin = 0,
ymax = 0.5,
width = 3.3in,
height = 1.75in,
grid = both,
xmin = -3,
xmax = 5,
xlabel = {$E_S / N_0$ (dB)},
ylabel = {Guess Count Reduction Proportion},
xtick={-3,-1,...,5},
ytick={0,0.1,...,0.5}
]

\addplot[blue, mark = +]
table{x y
-3 0.3637
-2.75 0.36455
-2.5 0.36577
-2.25 0.367
-2 0.36881
-1.75 0.36849
-1.5 0.37036
-1.25 0.37133
-1 0.37174
-0.75 0.37503
-0.5 0.37636
-0.25 0.37808
0 0.38007
0.25 0.38485
0.5 0.38789
0.75 0.38884
1 0.39221
1.25 0.39318
1.5 0.39287
1.75 0.39201
2 0.39237
2.25 0.38749
2.5 0.38408
2.75 0.36957
3 0.35738
3.25 0.33503
3.5 0.30878
3.75 0.27965
4 0.24832
4.25 0.21495
4.5 0.18422
4.75 0.15476
5 0.12741
};\addlegendentry{Relative \ac{SA-GCD} Savings}

\end{axis}
\end{tikzpicture}
\vspace{-24pt}
\caption{Relative guesswork savings of \ac{SA-GCD} is shown for the $(32, 21)$ SPC-aided \ac{BCH} code.  At each \ac{SNR} point, average guesswork savings of \ac{SA-GCD} is divided by average total guesswork of ORB-GCD.  \Ac{SA-GCD} achieves its greatest guesswork savings at \acp{SNR} where guesswork is most burdensome.}
\label{fig:overwork_eBCH_performance}

\end{figure}
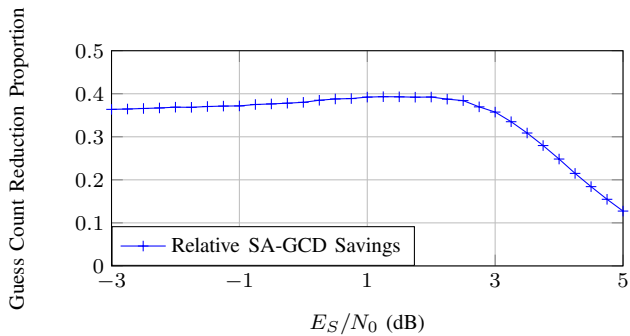

The $(32, 21)$ extended \ac{BCH} code was modified to support the SPC outer code structure required to realize complexity gains with \ac{SA-GCD}.  Rather than adding a parity bit to the $(31, 21)$ \ac{BCH} code to form the $(32, 21)$ e\ac{BCH} code, an SPC bit was added to the underlying \ac{BCH} code.  That is, an all-ones column was added to the generator matrix of the original \ac{BCH} code, producing a code to which we refer as a $(32, 21)$ SPC-aided \ac{BCH} code.  Results demonstrate that the SPC-aided \ac{BCH} code suffers mild distortion relative to the e\ac{BCH} code of matching dimensions, only $0.15$ dB in our measurements, see Figure \ref{fig:err_rate_eBCH_performance}.  Figures \ref{fig:guesswork_eBCH_performance} and \ref{fig:overwork_eBCH_performance} indicate that guesswork complexity is reduced as expected by the use of \ac{SA-GCD}.

\begin{figure}[t]
\centering
\footnotesize
\begin{tikzpicture}[scale=1]
\begin{semilogyaxis}[
legend style={at={(0, 0)},anchor= south west},
ymin=1e-5,
ymax=1,
width=3.3in,
height=2.25in,
grid=both,
xmin = -3,
xmax = 5,
xlabel = {$E_S / N_0$ (dB)},
ylabel = {BLER},
xtick={-3,-1,...,5}
]

\addplot[black, mark = x]
table{x y
-3 0.70292
-2.75 0.66027
-2.5 0.61407
-2.25 0.56528
-2 0.51423
-1.75 0.46165
-1.5 0.4086
-1.25 0.35603
-1 0.30497
-0.75 0.25658
-0.5 0.21112
-0.25 0.17011
0 0.13405
0.25 0.10285
0.5 0.076962
0.75 0.056111
1 0.039487
1.25 0.027022
1.5 0.017812
1.75 0.011366
2 0.0070673
2.25 0.0041601
2.5 0.0023661
2.75 0.0012936
3 0.00067237
3.25 0.00034305
3.5 0.00016097
3.75 7.05e-05
4 3.33e-05
4.25 1.35e-05
4.5 6e-06
4.75 2.2e-06
5 6e-07
};\addlegendentry{ORB-GCD, e\ac{BCH} Code}

\addplot[red, mark = o]
table{x y
-3 0.70622
-2.75 0.66434
-2.5 0.6195
-2.25 0.57125
-2 0.52091
-1.75 0.46829
-1.5 0.41615
-1.25 0.36302
-1 0.31297
-0.75 0.26374
-0.5 0.21829
-0.25 0.17664
0 0.13988
0.25 0.10833
0.5 0.082385
0.75 0.06033
1 0.043331
1.25 0.029848
1.5 0.020187
1.75 0.012971
2 0.008283
2.25 0.0050985
2.5 0.0029802
2.75 0.0016901
3 0.00093027
3.25 0.00047383
3.5 0.00023851
3.75 0.00012193
4 5.84e-05
4.25 2.54e-05
4.5 1.24e-05
4.75 4.9e-06
5 2.3e-06
};\addlegendentry{ORB-GCD, SPC-aided \ac{BCH} Code}

\addplot[blue, mark = +]
table{x y
-3 0.706
-2.75 0.66441
-2.5 0.61838
-2.25 0.56994
-2 0.51968
-1.75 0.46796
-1.5 0.41459
-1.25 0.36148
-1 0.31154
-0.75 0.26227
-0.5 0.2164
-0.25 0.17543
0 0.13904
0.25 0.10725
0.5 0.080589
0.75 0.059359
1 0.042299
1.25 0.029198
1.5 0.019686
1.75 0.012825
2 0.0080062
2.25 0.0048592
2.5 0.0028261
2.75 0.0016475
3 0.0008593
3.25 0.00045116
3.5 0.00024099
3.75 0.0001107
4 5.18e-05
4.25 2.28e-05
4.5 1.2e-05
4.75 4.3e-06
5 1.6e-06
};\addlegendentry{\Ac{SA-GCD}, SPC-aided \ac{BCH} Code}

\end{semilogyaxis}
\end{tikzpicture}
\vspace{-24pt}
\caption{\Ac{SA-GCD} achieves identical \ac{BLER} performance to ORB-GCD for the $(32, 21)$ SPC-aided \ac{BCH} code.  The SPC-aided \ac{BCH} code shows a slight distortion relative to the original e\ac{BCH} code, approximately 0.15 dB at high \acp{SNR}.  At low \acp{SNR} results were not statistically significant enough to yield a clear distortion value.}

\label{fig:err_rate_eBCH_performance}
\end{figure}
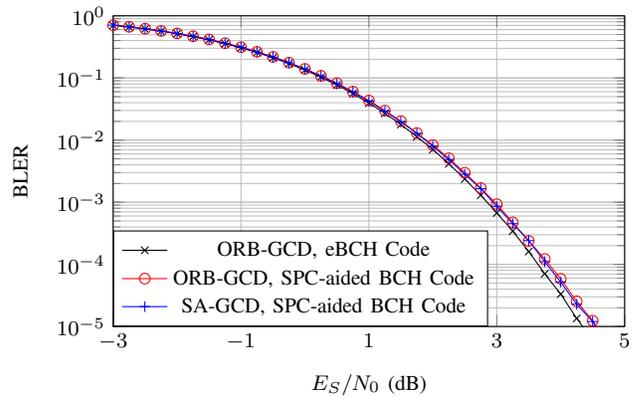

\begin{figure}[t]
\centering
\footnotesize
\begin{tikzpicture}[scale = 1]
\begin{semilogyaxis}[
legend style={at={(0, 0)}, anchor= south west},
ymin=1e-3,
ymax=1e1,
width=3.3in,
height=2.0in,
grid=both,
xmin = -3,
xmax = 5,
xlabel = {$E_S / N_0$ (dB)},
ylabel = {Average Guess Count per Message Bit},
ytick={0.001, 0.01, 0.1, 1, 10},
xtick={-3,-1,...,5}
]

\addplot[blue, mark = +]
table{x y
-3 2.2259
-2.75 2.159
-2.5 2.0831
-2.25 1.9935
-2 1.9002
-1.75 1.7776
-1.5 1.6608
-1.25 1.5243
-1 1.3885
-0.75 1.2559
-0.5 1.1109
-0.25 0.97421
0 0.84045
0.25 0.72141
0.5 0.60719
0.75 0.49974
1 0.40899
1.25 0.32785
1.5 0.25841
1.75 0.20221
2 0.15937
2.25 0.12267
2.5 0.095267
2.75 0.072474
3 0.056373
3.25 0.043007
3.5 0.033095
3.75 0.025584
4 0.01988
4.25 0.01538
4.5 0.012048
4.75 0.0094156
5 0.0073205
};\addlegendentry{\Ac{SA-GCD} savings over basic ORB-GCD}

\end{semilogyaxis}

\end{tikzpicture}
\vspace{-24pt}
\caption{Average saved guesswork per bit for the $(32, 21)$ SPC-aided \ac{BCH} code is shown for \ac{SA-GCD}.  \Ac{SA-GCD}  executes less guesswork on average than ORB-GCD, so guesswork savings is always positive.  For all symbol \acp{SNR} with practical decoder error rates, \ac{SA-GCD} realizes greater guesswork savings as symbol \ac{SNR} decreases.}
\label{fig:guesswork_eBCH_performance}
\vspace{-18pt}
\end{figure}
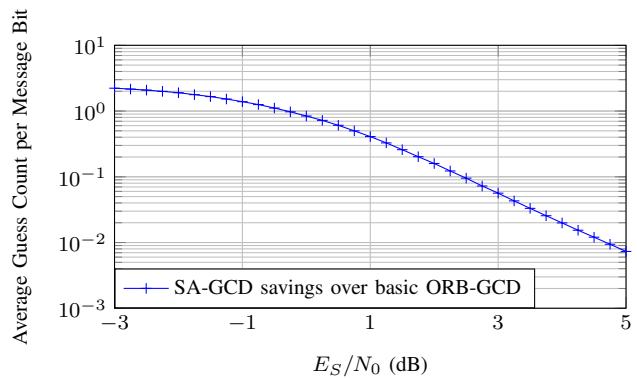

\begin{figure}[t]
\centering
\footnotesize
\begin{tikzpicture}[scale=1]
\begin{axis}[
legend style = {at = {(0, 0)}, anchor = south west},
ymin = 0,
ymax = 0.5,
width = 3.3in,
height = 1.75in,
grid = both,
xmin = -1,
xmax = 7,
xlabel = {$E_S / N_0$ (dB)},
ylabel = {Guess Count Reduction Proportion},
xtick={-1, 1,...,7},
ytick={0,0.1,...,0.5}
]

\addplot[blue, mark = +]
table{x y
-1 0.41731
-0.75 0.4124
-0.5 0.41517
-0.25 0.4128
0 0.4074
0.25 0.41842
0.5 0.40953
0.75 0.42244
1 0.41608
1.25 0.41354
1.5 0.42093
1.75 0.42618
2 0.44027
2.25 0.43054
2.5 0.42668
2.75 0.45131
3 0.44012
3.25 0.4435
3.5 0.43804
3.75 0.42007
4 0.4099
4.25 0.38705
4.5 0.36463
4.75 0.32345
5 0.27834
5.25 0.23155
5.5 0.18986
5.75 0.15089
6 0.11813
6.25 0.090952
6.5 0.068819
6.75 0.051409
7 0.037661
};\addlegendentry{Relative SA-GCD Savings}

\end{axis}
\end{tikzpicture}
\vspace{-24pt}
\caption{Relative guesswork savings of \ac{SA-GCD} is shown for the $(64, 48)$ SPC-aided \ac{CRC} code.  At each \ac{SNR} point, average guesswork savings of \ac{SA-GCD} is divided by average total guesswork of ORB-GCD.}
\label{fig:overwork_CRC}
\vspace{-18pt}
\end{figure}
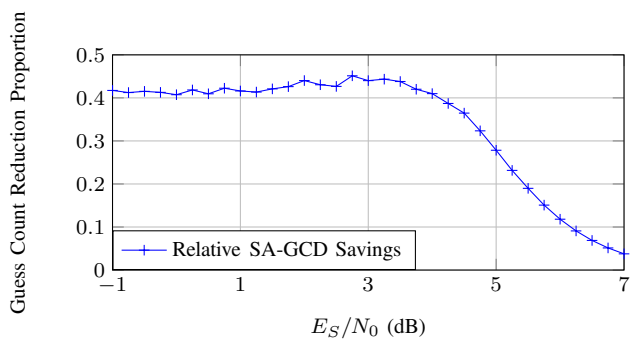

Simulations were also run to compare \ac{SA-GCD} performance on the $(64, 48)$ \ac{CRC} code generated by the polynomial $x^{16} + x^{12} + x^{5} + 1$ as used for downlink CA-Polar codes in the 5G standard \cite{3gpp.38.212}.  Figure \ref{fig:BLERs_CRC} indicates that swapping out a parity-check bit in the \ac{CRC} code for an SPC bit introduces little distortion, and Figures \ref{fig:guesswork_CRC} and \ref{fig:overwork_CRC} indicate the total and relative guesswork savings of \ac{SA-GCD} for this code.

\begin{figure}[t]
\centering
\footnotesize
\begin{tikzpicture}[scale=1]
\begin{semilogyaxis}[
legend style={at={(0, 0)},anchor= south west},
ymin=1e-5,
ymax=1,
width=3.3in,
height=2.25in,
grid=both,
xmin = -1,
xmax = 7,
xlabel = {$E_S / N_0$ (dB)},
ylabel = {BLER},
xtick={-1, 1, 3, 5, 7}
]

\addplot[black, mark = x]
table{x y
-1 0.73936
-0.75 0.67341
-0.5 0.60273
-0.25 0.52894
0 0.44599
0.25 0.37361
0.5 0.29851
0.75 0.23471
1 0.17459
1.25 0.12785
1.5 0.090153
1.75 0.06126
2 0.03999
2.25 0.025331
2.5 0.01545
2.75 0.0094969
3 0.0054109
3.25 0.0031344
3.5 0.0016777
3.75 0.00092773
4 0.00056011
4.25 0.00028128
4.5 0.00014073
4.75 7.14e-05
5 3.29e-05
5.25 1.57e-05
5.5 5.3e-06
5.75 3.1e-06
6 9e-07
6.25 2e-07
6.5 0
6.75 0
7 0
};\addlegendentry{ORB-GCD, no SPC bit}

\addplot[red, mark = o]
table{x y
-1 0.73692
-0.75 0.67093
-0.5 0.60166
-0.25 0.52551
0 0.44391
0.25 0.37097
0.5 0.30123
0.75 0.2331
1 0.17358
1.25 0.12554
1.5 0.08826
1.75 0.059033
2 0.038635
2.25 0.024337
2.5 0.01476
2.75 0.0090072
3 0.0051171
3.25 0.0029456
3.5 0.0016772
3.75 0.00087405
4 0.00050553
4.25 0.00026468
4.5 0.00014242
4.75 7e-05
5 3.49e-05
5.25 1.52e-05
5.5 7.3e-06
5.75 3.9e-06
6 2.2e-06
6.25 8e-07
6.5 2e-07
6.75 0
7 0
};\addlegendentry{ORB-GCD, with SPC bit}

\addplot[blue, mark = +]
table{x y
-1 0.73555
-0.75 0.67165
-0.5 0.60097
-0.25 0.52526
0 0.44886
0.25 0.37156
0.5 0.29838
0.75 0.23165
1 0.17179
1.25 0.12561
1.5 0.087873
1.75 0.058974
2 0.038441
2.25 0.024085
2.5 0.014812
2.75 0.0087696
3 0.0050859
3.25 0.0029106
3.5 0.0016768
3.75 0.00093243
4 0.00047695
4.25 0.00027448
4.5 0.00013473
4.75 7.1e-05
5 3.15e-05
5.25 1.72e-05
5.5 7.4e-06
5.75 2.5e-06
6 1.7e-06
6.25 7e-07
6.5 3e-07
6.75 0
7 0
};\addlegendentry{SA-GCD using SPC bit}

\end{semilogyaxis}
\end{tikzpicture}
\vspace{-24pt}
\caption{ORB-GCD with the $(64, 48)$ \ac{CRC} code suffers no more than $0.1$ dB of distortion with the replacement of a parity-check for an SPC bit, and the use of \ac{SA-GCD} causes no further distortion.}
\label{fig:BLERs_CRC}
\vspace{-18pt}
\end{figure}
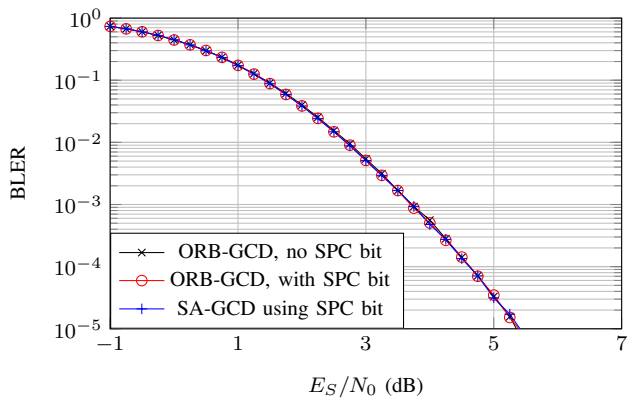

\begin{figure}[t]
\centering
\footnotesize
\begin{tikzpicture}[scale = 1]
\begin{semilogyaxis}[
legend style={at={(0, 0)}, anchor= south west},
ymin=1e-3,
ymax=1e2,
width=3.3in,
height=2.0in,
grid=both,
xmin = -1,
xmax = 7,
xlabel = {$E_S / N_0$ (dB)},
ylabel = {Average Guess Count per Message Bit},
ytick={0.001, 0.01, 0.1, 1, 10, 100},
xtick={-1,1,...,7}
]

\addplot[blue, mark = +]
table{x y
-1 36.169
-0.75 33.315
-0.5 30.713
-0.25 27.444
0 23.429
0.25 20.593
0.5 16.478
0.75 13.629
1 10.179
1.25 7.5553
1.5 5.5008
1.75 3.8694
2 2.7215
2.25 1.6903
2.5 1.0444
2.75 0.69368
3 0.40569
3.25 0.2484
3.5 0.14904
3.75 0.087847
4 0.054914
4.25 0.035219
4.5 0.023755
4.75 0.015854
5 0.010926
5.25 0.0076359
5.5 0.0055177
5.75 0.0039901
6 0.0029195
6.25 0.0021397
6.5 0.0015632
6.75 0.0011387
7 0.0008191
};\addlegendentry{SA-GCD savings over basic ORB-GCD}

\end{semilogyaxis}

\end{tikzpicture}
\vspace{-24pt}
\caption{Average saved guesswork per bit for the $(64, 48)$ SPC-aided \ac{CRC} code is shown for \ac{SA-GCD}.  \Ac{SA-GCD}  executes less guesswork on average than ORB-GCD, so guesswork savings is always positive.  For all symbol \acp{SNR} with practical decoder error rates, \ac{SA-GCD} realizes greater guesswork savings as symbol \ac{SNR} decreases.}
\label{fig:guesswork_CRC}
\end{figure}
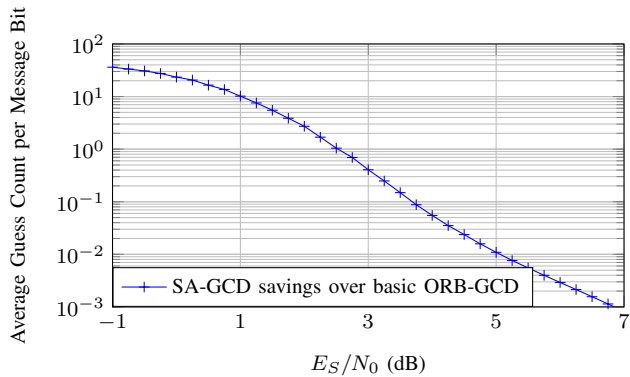

The highest guesswork percentage is saved at SNRs in the middle of each code’s waterfall curve.  Decoding involves greater guesswork in this \ac{SNR} regime than at high \acp{SNR}, but the decoder output is still highly reliable.  \Ac{SA-GCD} provides the greatest total guesswork savings at low \acp{SNR}, where guesswork is most burdensome.  At very low \ac{SNR}, the decoder output becomes unreliable because the code is not powerful enough.  At \acp{SNR} with more practical \acp{BLER} around $10^{-3}$, \ac{SA-GCD} realizes a larger proportional guesswork savings of a smaller guesswork task, so total guesswork is reduced less at middling \acp{SNR}.

\section{Conclusion}
\label{section:conclusion}
If the code in use can be structured as the concatenation of an SPC outer code with an arbitrary inner code, \ac{SA-GCD} offers a no-cost improvement over ORB-GCD.  The \ac{BLER} performance does not measurably change, and the guess count can be reduced by a factor of up to 2.  Modifying codes without the required structure is straightforward.  In the case of e\ac{BCH} codes and \ac{CRC} codes, replacing a redundancy bit with an SPC bit to enable the use of \ac{SA-GCD} causes limited distortion.

\section*{Acknowledgment}
This work was supported by the Defense Advanced Research Projects Agency under Grant HR00112120008.

\bibliographystyle{IEEEtran}
\bibliography{bibtex/main}

\end{document}